\documentclass[manuscript,screen]{acmart} 
\usepackage{tabularx} 
\usepackage{multirow} 
\AtBeginDocument{%
  }

\setcopyright{none} 
\acmDOI{XXXXXXX.XXXXXXX} 
\begin{document}

\title{Vulnerability Disclosure or Notification? 
Best Practices for Reaching Stakeholders at Scale}


\author{Ting-Han Chen} 
\authornote{Both authors contributed equally to this research.}
\email{t.h.chen@utwente.nl}
\orcid{0000-0001-8759-6676}
\author{Jeroen van der Ham-de Vos}
\authornotemark[1]
\email{j.vanderham@utwente.nl}
\affiliation{%
  \institution{University of Twente}
  \city{Enschede}
  \state{Overijssel}
  \country{The Netherlands}
}









\begin{abstract}
Security researchers are interested in security vulnerabilities, but these security vulnerabilities create risks for stakeholders. Coordinated Vulnerability Disclosure has been an accepted best practice for many years in disclosing newly discovered vulnerabilities. This practice has mostly worked, but it can become challenging when there are many different parties involved. 

There has also been research into known vulnerabilities, using datasets or active scans to discover how many machines are still vulnerable. The ethical guidelines suggest that researchers also make an effort to notify the owners of these machines. We identify that this differs from vulnerability disclosure, but rather the practice of vulnerability notification. This practice has some similarities with vulnerability disclosure but should be distinguished from it, providing other challenges and requiring a different approach.

Based on our earlier disclosure experience and on prior work documenting their disclosure and notification operations, we provide a meta-review on vulnerability disclosure and notification to observe the shifts in strategies in recent years. We assess how researchers initiated their messaging and examine the outcomes. We then compile the best practices for the existing disclosure guidelines and for notification operations.
\end{abstract}




\maketitle

\section{Introduction}
Coordinated Vulnerability Disclosure (CVD), formerly known as responsible disclosure, is the best practice accepted in the security community for dealing with discovered vulnerabilities. When a new vulnerability is found, the finder conducts a risk and impact assessment and then initiates conversations with stakeholders to mitigate vulnerable systems or services in time. 
However, due to the shift in the network landscape, the scale of vulnerable systems and stakeholders involved has changed drastically from the past~\cite{max_cvd_2023,durumeric_heartbleed_2014}. The amount and variety of vulnerabilities have increased~\cite{nakajima_pilot_2019}, and so have the affected parties~\cite{boucher_trojan_2022,li_youve_2016}. This presents a new challenge for finders, particularly academic security researchers and practitioners: figuring out a scalable method to identify reliable contact information and ensure message delivery to stakeholders.


Over the years, support mechanisms, such as vulnerability disclosure policy \cite{reidsma_operationalizing_2023}, VINCE vulnerability platform \cite{vince}, and bug bounty programs~\cite{walshe_bugbounty_2020}, have come into place to handle the challenge in vulnerability disclosure. Finders and vendors can adopt CVD to perform disclosure with the support organisations such as Computer Security Incident Response Teams (CSIRTs) \cite{CSIRTServicesFramework} and Product Security Incident Response Teams (PSIRTs) \cite{PSIRTServicesFramework}. However, not every finder or vendor has the same capacity and experience to conduct disclosure at scale \cite{walshe_coordinated_2022}. Moreover, finders from different communities, such as academic security researchers \cite{boucher_trojan_2022}, individual ethical hackers, practitioners, and bug bounty hunters, may have different security interests and constraints in selecting their approaches to reach stakeholders. 

Despite the challenge in vulnerability disclosure, notifications to stakeholders about known vulnerabilities and security issues that remain in existing systems have gained attention over the years~\cite{max_cvd_2023}. We identify the practice as vulnerability notification that informs end-users, such as product owners, hosting providers, domain owners, network operators, and incident responders. These stakeholders are different from vendors, who are traditionally the stakeholders in CVD. Whether or not the vulnerability is possibly known to the stakeholders and the public is the difference between vulnerability disclosure and vulnerability notification. In vulnerability \textit{disclosure}, a finder discovers a new vulnerability and plans to disclose it to the vendor. Meanwhile, in vulnerability \textit{notification}, a finder locates a known vulnerability on existing machines and notifies the end-users. This means that vulnerability notification is often conducted after the vulnerability disclosure. In particular, after a newly found vulnerability has been disclosed and mitigated with stakeholders or even made public, it is not guaranteed that all vulnerable systems are treated on a timely basis~\cite{durumeric_heartbleed_2014}. Hence, vulnerability notification is there to improve the safety of the internet. 

Vulnerability notification is often initiated after a finder carries out network scanning~\cite{vanderham_measurements_2017} or vulnerability analysis on existing datasets~\cite{chen_cvd_2024} to locate the vulnerable systems with targeted vulnerabilities. A finder should retrieve the stakeholders' contact information and select a communication channel to inform the affected parties. However, similar to the challenge in vulnerability disclosure, notification to stakeholders also suffers from the high number and complexity of stakeholders~\cite{stivala_pdf_2024,cetin_make_2017}. The contact retrieval and notification at scale pose an even tougher challenge to finders~\cite{maass_letter_2021,stock_hey_2016} since stakeholders in vulnerability notification may have a more diverse and complex set of parties involved~\cite{cetin_make_2017,chen_cvd_2024}.

\subsection{Research Questions}
\label{sec:rq}
How to disclose a vulnerability to a vendor is well-considered, with CVD as the best practice and increasing support mechanisms suggested to finders, governments and vendors~\cite{oecd_product_2021}. 
However, it is still unclear how well the best practices and mechanisms are known in the academic security research community. Furthermore, the lack of best practices for vulnerability notification has posed a growing challenge to not only academic security researchers but also other stakeholders. This motivates us to compile the research questions as follows:

\begin{enumerate}
    \item What distinguishes vulnerability disclosure and notification, especially in communication to stakeholders?
    \item How did academic researchers adopt best practices to carry out vulnerability disclosure and notification at scale?
    \item What insights can we gain from the academic researchers' experiences of large-scale disclosure and notification?
    \item What are the best practices for researchers and other finders to perform vulnerability disclosure and notification?
\end{enumerate}

In the following sections, we aim to answer each question with the insights we gain from literature, community, and support organisations, as well as our own disclosure and notification experience. In Section \ref{sec:background}, we distinguish between vulnerability disclosure and notification, and explain how a large-scale scenario can affect the two distinct practices in operations. In Section \ref{sec:selection}, we explain how we select literature and perform a meta-review with our proposed stage model to extract experiences from the academic security research community. In Section \ref{sec:lessons}, we compile the insights we gather from the academic experience in each stage of vulnerability disclosure and notification. Finally, in Section \ref{sec:bestpractices}, we propose our best practices for the finders to address the current limitations of large-scale vulnerability disclosure and notification. The best practices can also help other stakeholders to improve their disclosure and notification handling.

\section{Background and Related Work}
\label{sec:background}

\subsection{Vulnerability Disclosure and Notification}
\label{sec:disclosure_difference}

\begin{figure}
    \centering
    \includegraphics[width=1\linewidth]{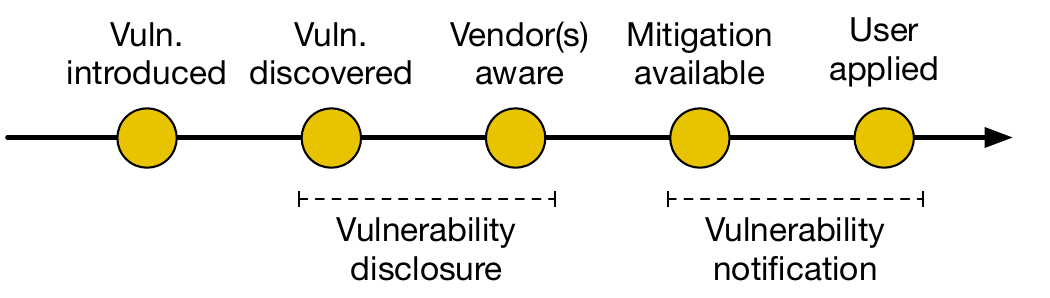}
    \caption{Vulnerability Disclosure and Notification}
    \label{fig:VD-ND}
\end{figure}

Vulnerability disclosure, especially coordinated vulnerability disclosure, has been an accepted best practice in security research for years. There have been numerous publications and even standards published on this process~\cite{oecd_product_2021}. However, notifying users with vulnerable systems has usually not been seen as a separate practice, but can come with challenges of its own. The difference between the two processes is illustrated in Figure~\ref{fig:VD-ND}.

In \textit{vulnerability disclosure}, a finder discovers a new vulnerability and aims to share this with the vendor. The role of the finder in the following content includes both the responsibilities of identifying a vulnerability and informing the responsible parties~\cite{first_guideline_2020}. Once the disclosure process has started, the vulnerability is usually revealed for the first time. The finder and vendor(s) discuss the finding with or without a coordinator's support. Subsequently, the vendor works on a mitigation.
Vulnerability disclosure requires trusted channels and detailed message information to minimise disclosure leaks to unintended recipients. The finder may need to expect extra steps to communicate with the stakeholder on whether or not and how to disclose the vulnerabilities to the public, based on the disclosure policy and legal agreement on both ends. After some time, the vulnerability is made public, usually with a possible mitigation available.

In \textit{vulnerability notification}, the vulnerability information is already available to the stakeholders before a notification starts. This means the vulnerability has already been documented and revealed to the stakeholders or the public. The weakness is likely documented with CVE numbers or discussed within specific communities. Potential malicious parties, such as criminals, may also be aware of the vulnerabilities and attempt to exploit them. 
The receiving users may or may not be aware of the vulnerability's existence before seeing the notification message. Once they are aware of the issues, they may be able to find the vulnerability information with resources that are not limited to the finder. This can lead to different user response behaviours to the finder. The affected user may also have a different risk assessment based on the severity of the vulnerabilities. The aim of the notification process is to improve general security by removing vulnerable systems.

We note that there are many overlapping challenges in vulnerability disclosure and notification, such as formulating the initial message, choosing the communication channel, and dealing with responses. However, due to the nature of the different stakeholders involved and especially the difference in scale, the challenges are fundamentally different.

\subsection{Evaluations of Disclosure and Notification}
\label{sec:evaluations}

Vulnerability disclosure has received best practices over the years through the contributions from security researchers, vendors, and support organisations. There have been several publications describing the practices in detail and examining their effectiveness. 
Householder and Spring \cite{householder_skillful_2022} proposed a model to assess the coordinator role in CVD, which provides insight into incident handling of vendors and the vulnerability lifecycle. Walshe and Simpson \cite{walshe_coordinated_2022} examined the effectiveness of how vendors with CVD programs and outsourced vulnerability platforms receive and process vulnerability disclosure. They revealed that disclosure program operators face a large number of vulnerability reports, which becomes a burden in vulnerability prioritisation. Nakajima et al.~\cite{nakajima_pilot_2019} looked into vulnerability management of IoT vendors across two countries and pointed out the disclosure pitfalls to avoid.

Vulnerability notification has received growing attention over the last decade, especially in the review of the notification efficiency. 
The Dutch Institute for Vulnerability Disclosure (DIVD) \cite{divd} proposed the notification guideline~\cite{max_cvd_2023} based on their framework and the Communication-Human Information Processing (C-HIP) model~\cite{wogalter_chip_1999} with their notification operation experiences. Their contribution focuses on scalable notification strategies for end-users, such as an incident responder and an abuse specialist. Additionally, the challenges they documented with end-users resemble the capacity and awareness issues of program operators discussed by Walshe and Simpson in vulnerability disclosure\cite{walshe_coordinated_2022}. 




Evaluations of vulnerability disclosure and notification at scale still require attention despite contributions trying to address the challenges from different aspects over the years. An extensive assessment and best practices for performing disclosure, especially notification at scale, are needed. 
There has been research focusing on finders trying to reach out to other stakeholders effectively and documenting the complete process of their disclosure and notification operations. The work will be discussed in the following sections with our assessment and comparison with best practices suggested by the communities and support organisations.


\subsection{Large-Scale Disclosure and Notification}
\label{sec:large-scale}

Large-scale disclosure operations have been reported as an increasing challenge by studies in the last decade~\cite{householder_skillful_2022, walshe_coordinated_2022, nakajima_pilot_2019}. Unlike the one-to-one or one-to-multiple disclosure to vendors, the complexity and number of stakeholders to inform have increased to a level beyond the effort individuals and small teams can make. 

Many of the challenges associated with multi-party disclosure processes have been identified by CERT/CC \cite{householder_cert_2017} and the FIRST Special Interest Group on Vulnerability Coordination \cite{VulnerabilityCoordinationSIG}, resulting in a best practice document "Guidelines and practices for Multi-Party Vulnerability Coordination and Disclosure" \cite{first_guideline_2020}. These best practices address the challenges of disclosing newly discovered vulnerabilities and, through various scenarios, describe the challenges and impacts for different stakeholders, including finders, vendors, defenders, and users. The VINCE platform~\cite{vince} has been developed by CERT/CC and other contributors to support multi-party disclosures and prevent many of the possible mistakes identified in the best practice guide.

Moreover, many countries have introduced laws, regulations \cite{cra}, and vulnerability disclosure policies \cite{BOD_CISA, ncsc_guideline_2018} to stimulate stakeholders, with active support from national CSIRTs \cite{CSIRTServicesFramework} and PSIRTs \cite{PSIRTServicesFramework}, professional organisations \cite{oecd_product_2021, first_guideline_2020}, and others, all supporting the practice of vulnerability disclosure. Likewise, research institutes, universities, and companies, such as Google and Facebook, have adopted outgoing disclosure policies to protect the finders and make their intentions clear to receiving parties.~\cite{reidsma_operationalizing_2023, max_cvd_2023}.

A vulnerability disclosure to multiple parties is considered to be large-scale when there are five or more vendors involved, and can become very complicated and stressful~\cite{koen_rpkiller_2023, boucher_trojan_2022, moura_stressful_2023}. However, vulnerability \textit{notification} processes can quickly become large-scale, and the affected parties can be counted in the hundreds or even thousands. Identifying the contact information and notifying the responsible parties behind the vulnerable systems can be overwhelming for finders~\cite{stock_hear_2018, cetin_make_2017, maass_letter_2021, stivala_pdf_2024}, which can eventually limit the development of vulnerability notification. 

There have been contributions to improve the efficiency of the notification mechanism. Emails have been empirically confirmed as the method of reaching large numbers of affected parties despite the drawback of low delivery rate and inaccurate contact information~\cite{stock_hey_2016, stock_hear_2018, cetin_make_2017, max_cvd_2023}. 

Nonetheless, the current best practices for vulnerability disclosure, especially vulnerability notification, are still struggling to catch up with the growing nature and importance of digital infrastructure. With best practices and prior experiences as a foundation for large-scale disclosure, the recent effort still suffers from the slow adoption of disclosure policy and incident handling among stakeholders \cite{koen_rpkiller_2023,chen_cvd_2024}.
Although bug bounty programs, vulnerability disclosure policies, and regulations are adopted as best practices by the industry and countries, not every finder from different communities, such as academic security researcher, practitioner, ethical hacker, and bounty hunter, shares the same capacity and security interest to ensure the message delivery and conversations of the disclosure and notification at scale with the current best practices and guidelines \cite{walshe_bugbounty_2020,boucher_trojan_2022,moura_stressful_2023}. For instance, academic security researchers may have time constraints for the publication schedule with their findings, while practitioners in a small team may have limited capacity and experience to prioritise stakeholders to inform when large-scale scenarios are considered. The current best practices do not necessarily cover the disclosure or notification guidelines for such finders.

There are increasingly large-scale internet-wide vulnerability notification cases beyond the support of local or national support organisations, which lead to finders, such as security researchers, carrying out the notifications by setting up a messaging infrastructure themselves~\cite{stivala_pdf_2024, chen_cvd_2024}. It has been reported by finders that contact identification prioritisation remains challenging, and the notification operation is still not broadly accepted among stakeholders~\cite{moura_stressful_2023,chen_cvd_2024}, even though disclosure policies or vulnerability report programs were present by stakeholders. This indicates that large-scale notifications still necessitate new proposals to enhance the process for both notifying and receiving parties. 

So far, we have answered our first research question in Section \ref{sec:rq}.
With a clear distinction between vulnerability disclosure and notification at scale, we aim to fill the gap by performing a meta-review on the experiences of academic security researchers, along with suggestions from other communities, support organisations, and our prior experience.

\section{Literature Selection}
\label{sec:selection}

To understand how the academic security research community approaches vulnerability disclosure and notification at scale over time, we collected publications with extensive documentation on large-scale vulnerability disclosure or notification operations in the last decade. We initially used literature search engines to find publications and then performed a selection based on the inclusion of disclosure procedures in stages and documentation before and after disclosure or notification operations. We did not exhaustively locate all available disclosure or notification work. Rather, we selected work that can be representative of the large-scale disclosure and notification implementation in different disclosure and notification scenarios from the academic security researcher's perspective.

The initial search patterns we used were the combination of "vulnerability", "disclosure", "notification",  "report", "large-scale", and "network" to look for the publications through major academic security conferences, literature databases, and search engines such as ACM Digital Library, IEEE Xplore, Scopus, and Google Scholar with all fields or metadata search filter enabled. We skimmed through the title, abstract, and partial content of search results that fit into the large-scale disclosure or notification scenarios discussed in Section \ref{sec:large-scale}. The initial result was broad and diverse. After a few iterations, we narrowed down our search keywords to "vulnerability disclosure" and "vulnerability notification", with synonyms such as "vulnerability alert" and "vulnerability warning".  It is worth noting that we also examine publications that don't necessarily have such keywords but have related content with manual effort. This gave us a short list of approximately 60 results that were relevant to vulnerability disclosure and notification.

To finalise the short list of publications, we examined the literature candidates with details such as assessment before an operation, selection of communication channels and messaging infrastructure, review after the operations, and contribution to best practices. We assigned different categories based on communication channels and involved stakeholders to identify areas of difference and choose work that has a measurable impact on the stakeholders. Furthermore, to understand the development of disclosure and notification best practices over the years, we aimed for publications that can be used to compare with each other, such as adopting, changing or improving methods based on other selected work. This means we looked into the reference list and related work sections of the publications to seek correlations and influences. Besides, we also selected work that made an impact in the security communities to illustrate the CVD as best practice and to stimulate the discussion on how we can improve the existing guidelines. Finally, we separated the publications into vulnerability disclosure and notification operations as mentioned in Section \ref{sec:disclosure_difference}.

As a result, we selected 15 distinct publications that well represent large-scale vulnerability disclosure and notification using several approaches in different scenarios from the last decade. We separate the work into two tables in chronological order. Table \ref{tab:disclosure} presents the vulnerability disclosure operations, whereas Table \ref{tab:notification} presents the vulnerability notification operations. To comprehend the selected work, we develop assessment stages to look into each disclosure or notification operation in the following subsection.



\subsection{Stages of Operation}
\label{sec:stage}

In the literature selection, we cross-reviewed each selected work in detail and aimed to figure out the common procedure for performing vulnerability disclosure and notification. Eventually, we compiled five stages to represent the procedure implemented across the literature. We then used the stages to extract key points, numbers, and remarks from the selected work to better understand efforts, considerations and reviews made in the operations. This results in the following stages to assess the selected publications:

\textbf{Pre-Assessment} -- Before a vulnerability disclosure or notification can start, a finder will assess the risk and impact of the discovered vulnerability. We identify the type and number of stakeholders involved and the vulnerabilities, and then extract the impact scale of the vulnerability to understand the preparation required before an operation.

\textbf{Communication Channel} -- After deciding on the stakeholders to inform, the proper communication channels should be selected to deliver the messages. We list the single to multiple communication channels used in each operation to inform the affected parties.

\textbf{Messaging Infrastructure} -- For the message to be delivered with the selected channel, the right messaging infrastructure should be used to ensure message delivery. We distinguish the messaging infrastructure used by finders to deliver the message, as well as the infrastructure used by the stakeholders to receive and forward the message. 

\textbf{Disclosure Policy and Message} -- The wording of the message should be tuned to the stakeholders to ensure comprehension and follow the needs of the recipients. We reflect on how the finders composed their message and handled the conversations with other stakeholders. In addition, we also check on disclosure policies used by the finders and other stakeholders, if presented.

\textbf{Post-Review} -- An operation can be reviewed by tracking the remediation rate, feedback from stakeholders, and experiences in each stage. A finder can also reflect on the operation and contribute to best practices. We extract the notified hosts and domains, remediation rates, interactions between the stakeholders, and contributions to best practices of large-scale disclosure and notification operations.

To present an overview of the assessment of each publication, we arrange three tables based on the above five stages. Tables \ref{tab:disclosure} and \ref{tab:notification} both contain the publication titles, pre-assessment, and post-review to indicate the efforts and outcomes from each work with different scenarios considered. Table \ref{tab:channel} shows the single or multiple communication channels and messaging infrastructure implemented in each reviewed disclosure or notification operation.

The five stages and the three tables serve as our response to the second research question presented in Section \ref{sec:rq}, as well as the foundation to understand the selected publications and extract the insights in the next section.

\renewcommand\tabularxcolumn[1]{m{#1}}

\begin{table*}
    \setlength{\extrarowheight}{1.1pt}
    \centering
    \begin{tabularx}{\linewidth}{XcXX}
    \hline 
    Title & Year & Pre-Assessment & Post-Review\\
    \hline 
    Key Reinstallation Attacks: \newline Forcing Nonce Reuse in WPA2~\cite{vanhoef_key_2017} & 2017 & To Wi-Fi related vendors, \newline Key reinstallation attacks, \newline At least 6 vendors & Increased to 84 vendors (VINCE), \newline Measurement and notification delegated to CERT/CC and vendors \\
    \hline     
    Talking Trojan: \newline Analysing an Industry-Wide Disclosure~\cite{boucher_trojan_2022} & 2022 & To software vendors, \newline Trojan Source, \newline 13 vendors & Increased to 19 vendors (VINCE),\newline Measurement delegated to vendors, GitHub, and Rust team \\ \hline 
    rpkiller: Threat Analysis of the BGP Resource Public Key Infrastructure~\cite{koen_rpkiller_2023} & 2023 & To software vendors, \newline RPKI implementations, \newline 8+ software vendors & 5 vendors released fixes, \newline 2 vendors didn't release the fix, \newline 1 vendor stopped support \\
    \hline  
    Vulnerability Disclosure \newline Considered Stressful~\cite{moura_stressful_2023} & 2023 & To vendors, network operators \newline TsuNAME (DNS resolver\&clients), \newline 5+ vendors, operator communities & Measurement not specified, \newline 1 DNS community meeting, \newline 3 security events\\
    \hline 
    Haunted by Legacy: \newline Discovering and Exploiting Vulnerable Tunnelling Hosts~\cite{beitis_haunted_2025} & 2025 & To vendors, domain owners, \newline Tunnelling protocol \& EDoS, \newline 3,527,565 IPv4, 735,628 IPv6 hosts & Up to 14 domains directly notified, \newline Measurement \& notification delegated to CSIRTs and Shadowserver \\
    \end{tabularx}
    \caption{Vulnerability Disclosure Operations}
    \label{tab:disclosure}
\end{table*}

\begin{table*}
    \setlength{\extrarowheight}{1.1pt}
    \centering
    \begin{tabularx}{\linewidth}{XcXX}
    \hline 
    Title & Year & Pre-Assessment & Post-Review \\
    \hline 
    The Matter of Heartbleed~\cite{durumeric_heartbleed_2014} & 2014 & To network operators, \newline TLS Heartbeat Extension, \newline 588,686 vulnerable hosts & 212,805 hosts notified, \newline 4,648 emails (WHOIS), \newline 57\% mitigated \\
    \hline
    You've Got Vulnerability: \newline Exploring Effective Vulnerability Notifications~\cite{li_youve_2016} & 2016 & To network operators, \newline 45,770 ICS, 83,846 DDoS Ampl., \newline 180,611 IPv6 Firewall hosts & 79.7\% ICS,92.4\% Ampl.,99.8\% IPv6 \newline notified, 9,918 emails (WHOIS), \newline Up to 18\% mitigated \\ 
    \hline
    Hey, You Have a Problem: On the Feasibility of Large-Scale Web Vulnerability Notification~\cite{stock_hey_2016} & 2016 & To website owners \newline WordPress(WP) \& Client-Side XSS, \newline 44,790 vulnerable domains & 35,832 domains notified, \newline  17,819 emails (alias, WHOIS), \newline 25.8\% WP, 12.6\% CXSS mitigated \\
    \hline   
    Make Notifications Great Again: Learning How to Notify in the Age of Large-Scale Vulnerability Scanning\cite{cetin_make_2017} & 2017 &  To nameserver/network operators, \newline DNS Dynamic Updates, \newline 21,506 vulnerable domains & 4,149 nameserver IPs notified, \newline 5051 emails (WHOIS), \newline Up to 20\% mitigated \\
    \hline
    Didn't You Hear Me? - \newline Towards More Successful Web Vulnerability Notifications~\cite{stock_hear_2018} & 2018 & To website owners, \newline WordPress(WP) \& Git, \newline 24,000+ vulnerable domains &  20,602 domains notified, \newline 103,819 emails (alias, WHOIS), \newline 17\% WP, 24\% Git  mitigated \\ 
    \hline
    Tell Me You Fixed It: Evaluating Vulnerability Notifications via Quarantine Networks~\cite{cetin_tell_2019} & 2019 & To ISP retail customers, \newline DNS resolvers \& mDNS services, \newline 1688 retail customers & 688 customers notified, \newline 86.6\% walled garden only, \newline 75.1\% email only mitigated \\
    \hline
    User Compliance and Remediation Success after IoT Malware Notifications~\cite{rodriguez_iot_2021} & 2021 & To ISP retail customers, \newline IoT Malware (Mirai Family), \newline 177 retail customers & From 50 responded participants, \newline 95\% walled garden, 82\% email only \newline started mitigation \\ 
    \hline
    Effective Notification Campaigns on the Web: A Matter of Trust, Framing, and Support~\cite{maass_letter_2021} & 2021 & To website owners, \newline IP Anonymisation Misconfigured, \newline 7979 non-compliant sites & 4594 owners of 4754 sites notified,\newline 2660 letters, 1,337 emails (Manual), \newline 76.3\% letter, 33.9\% email mitigated \\
    \hline    
    Uncovering the Role of Support Infrastructure in Clickbait PDF Campaigns~\cite{stivala_pdf_2024} & 2024 & To website owners, \newline Clickbait PDF, \newline 177,835 vulnerable hosts & 8,842 domains notified, \newline 1,522 emails (alias, WHOIS), \newline 29.567\% mitigated \\ 
    \hline 
    Are You Sure You Want To Do \newline Coordinated Vulnerability Disclosure?~\cite{chen_cvd_2024} & 2024 & To IoT backend operators, \newline MQTT \& Misconfigured Backends, \newline 15,820 vulnerable backends & 15046 backends notified, \newline 2,132 emails (WHOIS), \newline 2.25\% mitigated \\
    \end{tabularx}
    \caption{Vulnerability Notification Operations}
    \label{tab:notification}
\end{table*}


\section{Lessons Learned}
\label{sec:lessons}

This section presents the insights we have gained from the selected work on large-scale vulnerability disclosure and notification. The main focus is from the finder's perspective and academic research experience. We also integrate the experience we learn from the communities, support organisations, existing guidelines and standards to present the essence of the disclosure or notification procedure.


With Tables \ref{tab:disclosure}, \ref{tab:notification}, and \ref{tab:channel} presented as an overview of disclosure and notification operations, we first discuss the development of best practices from the experiences of the selected work in Section \ref{sec:development}. Then, we compare how the finders adopted the past best practices and what they learned from their operations in Section \ref{sec:comparison}.



\subsection{The Development of Best Practices}
\label{sec:development}
Academic security researchers
have been seeking optimal methods and strategies as finders based on the disclosure and notification best practices presented in their time, as shown in Table \ref{tab:disclosure} and \ref{tab:notification}. We have observed changes in their communication channels, messaging infrastructure, contact retrieval, message formulation, and disclosure policies, all of which are influenced by the nature of the vulnerabilities found, notification scalability, and community recommendations.


The essential part of vulnerability disclosure and notification is ensuring messages reach the responsible parties and prompt them to action. As presented in Table \ref{tab:channel}, the finders used different communication channels and messaging infrastructure to ensure the delivery of the disclosure messages, either on their own or with support organisations, such as national CSIRTs and The Shadowserver Foundation. To maximise notification coverage, all work adopted multiple communication channels, and most of them delegated notifications to support organisations.

The Shadowserver Foundation is a non-profit organisation founded in 2004, driven by the vision of a secure, threat-free internet \cite{ShadowserverFoundation}. They do this by scanning the internet for known vulnerabilities, performing notifications, managing sinkholes for malware command-and-control infrastructure, and operating honeypots and honeyclients to monitor threat developments. Over the years, the team at The Shadowserver Foundation has created a large, trusted network with national CSIRTs, ISPs and others to ensure that they are able to reach as many stakeholders as possible. Anyone can sign up to receive threat reports about their domain or IP, provided they can prove ownership. More importantly, The Shadowserver Foundation can support researchers in performing notifications to stakeholders, as described by Beitis and Vanhoef \cite{beitis_haunted_2025,shadowserver_tunnel_2025}.

\begin{table}[h!]
    \centering
    \begin{tabular}{ccl}
         \multicolumn{2}{c}{Channel \& Infrastructure} &  Work  \\ \hline
         \multirow{3}{*}{Email} & \multicolumn{1}{c}{Individual Account} & \cite{stock_hey_2016,stock_hear_2018,vanhoef_key_2017,koen_rpkiller_2023,moura_stressful_2023,stivala_pdf_2024,beitis_haunted_2025} \\\cline{2-3}
                                & \multicolumn{1}{c}{Dedicated Account} & \cite{li_youve_2016,cetin_make_2017,cetin_tell_2019,rodriguez_iot_2021,chen_cvd_2024,maass_letter_2021} \\\cline{2-3}
                                & \multicolumn{1}{c}{Not Specified} & \cite{durumeric_heartbleed_2014,boucher_trojan_2022} \\ \hline
         \multicolumn{2}{l}{Coordinator Delegation} & \cite{li_youve_2016,stock_hey_2016,vanhoef_key_2017,boucher_trojan_2022,koen_rpkiller_2023,beitis_haunted_2025} \\ \hline
         \multicolumn{2}{l}{Website} & \cite{li_youve_2016,cetin_make_2017,vanhoef_key_2017,boucher_trojan_2022,chen_cvd_2024,beitis_haunted_2025}\\ \hline
         \multicolumn{2}{l}{Survey} & \cite{durumeric_heartbleed_2014,li_youve_2016,stock_hear_2018,cetin_make_2017}\\ \hline
         \multicolumn{2}{l}{Phone} & \cite{stock_hey_2016,rodriguez_iot_2021,maass_letter_2021} \\ \hline
         \multicolumn{2}{l}{ISP Intervention} & \cite{cetin_tell_2019,rodriguez_iot_2021} \\ \hline
         \multicolumn{2}{l}{Community \& Meeting} & \cite{koen_rpkiller_2023,moura_stressful_2023}\\ \hline
         \multicolumn{2}{l}{Post} & \cite{stock_hey_2016,maass_letter_2021} \\ \hline
         \multicolumn{2}{l}{Social Media} & \cite{stock_hey_2016} \\ \hline
         \multicolumn{2}{l}{The Shadowserver Foundation} & \cite{beitis_haunted_2025} \\ \hline
    \end{tabular}
    \caption{Channel and Messaging Infrastructure}
    \label{tab:channel}
\end{table}

The earlier vulnerability notification operations in~\cite{durumeric_heartbleed_2014,li_youve_2016} focused on email and coordinator delegation notifications, then others in~\cite{cetin_make_2017,stock_hear_2018,stock_hey_2016} extensively implemented diverse communication channels, compared the effectiveness of each empirically, and provided valuable suggestions to best practices on large-scale notifications. These operations served as the foundation for the large-scale notification best practices, especially in the academic security community.

The recent notification operations in~\cite{chen_cvd_2024,stivala_pdf_2024} adopted best practices from the prior work. The researchers narrowed down the channel selection and dived into the experiences with email notifications. On the contrary, Maass et al.~\cite{maass_letter_2021} decided to explore alternatives to the best practices with costly manual efforts on email, post, and phone calls. All their results reflected the prior best practices and helped improve notification best practices at scale. In all the above work, locating accurate contact information remains the major challenge regardless of the selected communication channels. On the other hand, finders in \cite{cetin_tell_2019,rodriguez_iot_2021} collaborated with a medium-sized ISP to explore the effectiveness of email and specialised ISP notification systems with different vulnerabilities, which serves as an example in notification with direct intervention from an ISP to their customers. 

Vulnerability disclosure to vendors has received promising best practices as implemented in ~\cite{vanhoef_key_2017,boucher_trojan_2022,koen_rpkiller_2023}. Boucher and Anderson \cite{boucher_trojan_2022} have provided well-structured details on CVE applications, CERT/CC VINCE notification, public press disclosure, and constraints in academic publication. However, despite the stakeholders presenting bug bounty programs and disclosure policies, they encountered slow responses and time constraints in their disclosure operations, as well as in academic publication scheduling. On the other hand, the researchers in the two disclosure operations~\cite{vanhoef_key_2017,koen_rpkiller_2023} tackled different software vulnerabilities but encountered the challenge of inviting the same stakeholder to participate in CVD. Last but not least, the two operations in \cite{moura_stressful_2023,beitis_haunted_2025} cover both vulnerability disclosure and notification. Their experiences revealed the long-lasting challenges of transitioning from disclosure to large-scale notification, as well as the need to explore newer communication channels and strategies to mitigate vulnerabilities in a timely manner.

\subsection{The Comparison of Different Scenarios}
\label{sec:comparison}
To understand how the finders in the selected work adopted and contributed to best practices, we describe and compare their experiences based on the operation stages introduced in Section \ref{sec:stage}. The comparison follows the order of the stages: Pre-Assessment, Communication Channel, Messaging Infrastructure, Disclosure Policy and Message, and Post-Review.


\subsubsection{Pre-Assessment}

Once a new vulnerability is found, the finder should assess the possible impact, affected parties, and the possible risk associated with that impact. This will determine the best approach to inform stakeholders.





All disclosure operations~\cite{vanhoef_key_2017,boucher_trojan_2022, moura_stressful_2023,beitis_haunted_2025} initiated CVD directly to affected vendors with newly found vulnerabilities and later delegated their disclosure to support organisations. Before their disclosure, most finders could identify several vendors with bug bounty~\cite{boucher_trojan_2022} or tracking program~\cite{moura_stressful_2023}, and a short list of direct disclosure contact information~\cite{vanhoef_key_2017,boucher_trojan_2022,koen_rpkiller_2023,beitis_haunted_2025,moura_stressful_2023}, which ranges from 5 to 14 affected parties due to limited capacity or testing environments. However, most finders faced challenges such as limited contact information of other stakeholders and the potential development to large-scale disclosure. They then consulted national CSIRTs, CERT/CC (VINCE), or NCSC-NL, except for the researchers in~\cite{moura_stressful_2023}. The impact scale later increased dramatically from the cooperation with support organisations, which is presented in the Post-Review in Table \ref{tab:disclosure}. Despite the lack of contact information and prioritisation in the pre-assessment, the finders could still mitigate the challenges with the support organisations.

The notification operations are vulnerability notifications to the stakeholders, with hosts or domains remaining vulnerable after public disclosure. The finders either leveraged existing datasets~\cite{chen_cvd_2024} or performed network scanning~\cite{vanderham_measurements_2017} to assess the impact scale of the existing vulnerabilities or other security issues. The large scale of the notification operations listed in Table \ref{tab:notification}, unlike disclosure operations in Table \ref{tab:disclosure}, has reached significantly higher numbers, ranging from ten to a hundred thousand. In the latest operation in \cite{beitis_haunted_2025}, the affected IPv4 hosts have reached a million at an internet-wide scale. On the other hand, the two notification operations in \cite {cetin_tell_2019, rodriguez_iot_2021} have fewer affected parties due to the partnership with a medium-sized ISP with its selected customers. In addition, most notifications included more than one vulnerability or security issue of the vulnerable hosts and domains. Hence, the efforts to estimate the exact number of affected parties increased as well. 

As a result, the main challenge of vulnerability notification is how to inform diverse stakeholders and handle multi-party scenarios at scale. Based on the initial estimation of the vulnerability characteristic, impact scale, and affected parties, the next step is to choose the most efficient communication channels and messaging infrastructure to inform stakeholders effectively.

\subsubsection{Communication Channel}

Selecting a communication channel remains critical~\cite{wogalter_chip_1999} to vulnerability disclosure and notification. Depending on the pre-assessment, a finder can select single to multiple communication channels to reach out to stakeholders at scale, as shown in Table \ref{tab:channel}. With the current guidelines for finders to perform large-scale disclosure or notification, there is still no simple decision on channel selection considering the tradeoffs between contact information retrieval and resource capacity, as discussed in all selected works.

\textbf{For vulnerability disclosure}, CVD as best practice and support mechanisms, such as bug bounty programs, disclosure programs, and disclosure policies, are provided in guidelines and gradually adopted by vendors~\cite{walshe_coordinated_2022}. However, improvements for different scenarios are still being made to best practices. Boucher and Anderson \cite{boucher_trojan_2022} identified their initial list of affected vendors through bug bounty programs, direct disclosure contacts, and outsourced vulnerability platforms. Although the researchers succeeded in most bounty programs, they encountered delayed responses or complications in follow-up conversations with outsourced platforms due to the latter's prioritisation of vulnerability reports~\cite{walshe_coordinated_2022}. They concluded that direct disclosure contacts or bounty programs of vendors might be more effective in extended discussions for their disclosure. 

In comparison, Moura and Heidemann \cite{moura_stressful_2023} experienced late answers and delayed mitigation schedules with a well-known vendor's direct contact and a bug tracking system. On the other hand, all the finders in \cite{vanhoef_key_2017, boucher_trojan_2022, koen_rpkiller_2023, beitis_haunted_2025} except for Moura and Heidemann \cite{moura_stressful_2023} delegated the notifications to coordinators such as CERT/CC with VINCE and national CSIRTs. Finally, finders in \cite{moura_stressful_2023,koen_rpkiller_2023} attended conferences and meetings to reach out to vendors or operators in related communities.

\textbf{For vulnerability notification}, finding an optimised channel to inform the stakeholders at scale turns out to be more challenging than vulnerability disclosure. The most significant difference is that the coordinator delegation to national CSIRTs has been reported as ineffective due to the large scale that went beyond the coverage of the support organisations, as reported in \cite{li_youve_2016,stock_hey_2016}. The finders in \cite{cetin_make_2017,stock_hear_2018,stock_hey_2016} exhausted multiple communication channels shown in Table~\ref{tab:channel} and empirically studied the effectiveness of each channel with the remediation rates by time and feedback from affected parties. In the three publications, they all concluded that email still remains the prominent channel for conducting large-scale notifications with the best coverage rate despite the obvious pitfalls, such as a high bounce rate, spam filter, and low awareness of recipients. 

Similarly, the same concerns were reported by finders in \cite{durumeric_heartbleed_2014,li_youve_2016}, which used email as their primary channel. In recent years, finders in~\cite{chen_cvd_2024,stivala_pdf_2024} adopted best practices derived from the previous work and narrowed their selection to email with a website as support. On the contrary, Maass et al.~\cite{maass_letter_2021} alternatively selected post and email to compare the effectiveness of both and provided phone support to notified parties. Given the considerable manual effort required for email and post contract information retrieval, they concluded that the post could be an effective but costly channel in their nation. Eventually, the finders in~\cite{chen_cvd_2024,stivala_pdf_2024,maass_letter_2021} all confirmed that the aforementioned issues of notifying stakeholders via email remain.

Across all the selected publications, email is the widely used communication channel in initial messages and discussions with vendors and end-users. The main challenges of email are still contact retrieval and low delivery rate. The finders scripted database queries using WHOIS either with a purchased database as in \cite{stock_hear_2018,stock_hey_2016} or with an online query service as in \cite{chen_cvd_2024}, to retrieve email contact information. Although the finders in~\cite{stock_hear_2018,chen_cvd_2024} mentioned RDAP as a potential alternative, none attempted it. However, Fernandez et al. ~\cite{fernandez_whois_2024} reported that RDAP has not caught up with the coverage of the existing WHOIS database despite the protocol being introduced to improve contact sharing. Moreover, Maass et al.~\cite{maass_letter_2021} confirmed that manual contact retrieval, at best, does not guarantee accurate contact information either. Aside from the contact retrieval issue, email is constantly challenged by high bounce rate~\cite{durumeric_heartbleed_2014,li_youve_2016}, spam filter~\cite{stock_hear_2018,chen_cvd_2024}, and low incentives of recipients to read messages~\cite{cetin_make_2017}. This eventually results in a low delivery rate to the stakeholders.

Nevertheless, a dedicated scalable notification system with a notification partner can be a solution. Finders in ~\cite{cetin_tell_2019,rodriguez_iot_2021} teamed up with an ISP and achieved better results with the ISP's walled-garden notifications than email. On the other hand, Beitis and Vanhoef \cite{beitis_haunted_2025} partnered with The Shadowserver Foundation, which actively measures the mitigation progress and notifies its subscribers at scale. Their notification is a new attempt at a communication channel, and the outcome seems promising.

\subsubsection{Messaging Infrastructure}

A finder will build their messaging infrastructure or delegate it to support organisations based on the selected communication channel. The common practice of outgoing email and website is using a registered individual or disclosure-specific account with the domain name of the finder's organisation to increase the delivery rate and trust of recipients, as implemented in \cite{li_youve_2016,stock_hey_2016,stock_hear_2018,chen_cvd_2024}. The email account selection of each selected publication is presented in Table \ref{tab:channel}.  The dedicated account practice is to address the email notification challenge of spam and phishing mitigation.
Due to the increasing spam and phishing messages, more and more measures are in place to prevent the delivery of these messages. In the early 2000s, it was possible to spin up a mail server and immediately send out an email notification to thousands of recipients. These days, there are many different standards associated with sending out emails (SPF, DKIM, DMARC, etc.), as mentioned in \cite{li_youve_2016}, and the reputation of the mail server is taken into account before an email is delivered.

Although a finder or its organisation can maintain their messaging infrastructure, it is also reported in~\cite{chen_cvd_2024} that the finders' email infrastructure is partially outsourced to a mail server, Microsoft Exchange. Their mail server introduces several restrictions, such as limited account control and email-sending rate, which prevent sending out large amounts of messages~\cite{exchange_limitation} efficiently. This makes it harder for finders to perform large-scale notification operations using email. This also resonates with the trends of vendors outsourcing vulnerability platforms reported in \cite{boucher_trojan_2022}, which limit the reachability from the sender to the responsible parties, the client of the outsourced platforms.

There are disclosure and notification operations delegating the messaging to national CSIRTs (such as NCSC-NL), CERT/CC with VINCE~\cite{vince}, and The Shadowserver Foundation. These are listed as Coordinator Delegation and The Shadowserver Foundation in the Table \ref{tab:channel}. While VINCE serves as a disclosure database with notification to vendors, as adopted in \cite{boucher_trojan_2022,vanhoef_key_2017}, national CSIRTs have their mailing list or website to inform vendors and end-users, which was mentioned in \cite{li_youve_2016,stock_hey_2016,koen_rpkiller_2023,beitis_haunted_2025}. In particular, The Shadowserver Foundation has become a newer notification delegation option with the organisation's own messaging infrastructure to inform the stakeholders, including vendors and end-users, such as ISPs, domain owners, and network operators~\cite{ShadowserverFoundation}. Besides, if a finder works with an organisation with dedicated communication channels, such as an ISP notifying its customers with its internal notification and management tool in \cite{cetin_tell_2019,rodriguez_iot_2021}, there can be a more efficient way to trigger action from the recipients. Furthermore, finders in \cite{li_youve_2016,cetin_make_2017,chen_cvd_2024} observed that stakeholders, such as domain name owners and cloud providers, used their notification systems to prompt their customers to take action.

\subsubsection{Disclosure Policy and Message}

The formulation of the disclosure message can determine the attention of the recipients and whether or not to respond and take action~\cite{wogalter_chip_1999}. The tradeoffs of length and details of content, such as remediation and security suggestions, are extensively discussed in \cite{cetin_make_2017,stock_hey_2016,stock_hear_2018,cetin_tell_2019,li_youve_2016,chen_cvd_2024,stivala_pdf_2024}, where the authors document not only their disclosure message templates but also the feedback from stakeholders.

Finders in \cite{cetin_make_2017,stock_hey_2016,stock_hear_2018} provided brief vulnerability information, affected systems, and disclosure purposes in their initial messages. The messages prompted recipients to either respond to the emails or visit a webpage for more information on vulnerabilities and remediations, using the automatically generated token from each message provided by the finders. This way, the finders could monitor the response rate with a dedicated web backend. Moreover, the finders can reduce the risk of information leaks in cases involving incorrect recipients. Furthermore, disclosure messages could be embedded with HTML content, such as the logo of the finder's organisations, to further check if a recipient loads the full message. However, due to the spam filter and recent email client loading mechanism, extensive embedded HTML content is no longer a suggested method confirmed in~\cite{stock_hear_2018, cetin_make_2017}. 
From the common experiences of earlier \cite{cetin_make_2017,stock_hear_2018,li_youve_2016} to recent publications \cite{stivala_pdf_2024, chen_cvd_2024}, the finders confirmed that plain text is the suggested way with less distraction and distinction from phishing messages.

In large-scale vulnerability notifications, providing information such as domain names, IP addresses, ports, and issues found in the vulnerable systems is recommended to help stakeholders investigate issues in time, as documented from the stakeholders' feedback in \cite{li_youve_2016,stock_hear_2018,cetin_make_2017,chen_cvd_2024}. However, receiving parties may have certain mail filters for incoming messages, which strictly limit the message length and attachments, as reported in \cite{chen_cvd_2024}. This could hinder the case of cloud providers or domain owners with a large number of vulnerable systems running for their clients. Such stakeholders may decline the messages containing longer vulnerable host lists or attachments. This eventually makes the decision on the message content more challenging for finders \cite{cetin_make_2017,chen_cvd_2024,stivala_pdf_2024}. 


Organisations present support for CVD by publishing a disclosure policy on their websites, using security.txt\cite{rfc9116}, or providing bounty or tracking programs~\cite{boucher_trojan_2022,moura_stressful_2023,vanhoef_krack_2017}. Boucher and Anderson \cite{boucher_trojan_2022} observed vendors outsourcing their programs to third-party platforms that reveal their own policies and prioritisation on vulnerability selections. They noted that the policies and preferences of the outsourced platforms may limit the incentive and direct message delivery to the responsible parties. In the vulnerability notification, Chen et al. \cite{chen_cvd_2024} observed that most of the stakeholders still do not provide security.txt or equivalent information in their responses, but privacy policies that do not necessarily help the disclosure process. The implementation of the disclosure policies among stakeholders still requires attention.

Google Project Zero \cite{ProjectZeroVulnerability} started with an outgoing vulnerability disclosure policy describing the timelines they would use in disclosing vulnerabilities. On the one hand, this has pressured vendors to work on mitigation and publish it within 90 days of the timeline. After several years, the 90-day deadline has become an accepted practice. Academic researchers have also started using outgoing vulnerability disclosure policies \cite{reidsma_operationalizing_2023}, which help build trust between stakeholders and coordinate disclosure operations. The outgoing policy has been implemented in \cite{koen_rpkiller_2023,chen_cvd_2024}, where the finders presented their intended procedures on vulnerability handling, notification frequency, public disclosure schedule, and more \cite{ncsc_guideline_2018,vanderham_cvd_2023} in their outgoing messages. This gives the receiving party a brief yet informative message for potential procedures and conversation. Other finders in \cite{cetin_make_2017,cetin_tell_2019,stock_hear_2018,moura_stressful_2023,stivala_pdf_2024} did not document a dedicated disclosure policy, yet did provide equivalent information and contact methods in their messages for coordinating or exempting from the notifications. As for the finders that chose coordinator delegation, ISP intervention, and The Shadowserver Foundation, they also followed and presented the disclosure policy from the support organizations as documented in  \cite{boucher_trojan_2022,cetin_tell_2019,koen_rpkiller_2023,moura_stressful_2023,rodriguez_iot_2021,stock_hear_2018,stock_hey_2016,vanhoef_key_2017,vanhoef_krack_2017}. These delegated parties often use existing relations with stakeholders and customers to create a trustworthy communication channel.

\subsubsection{Post-Review}
\label{sec:lessons-postreview}
The remediation rate of the vulnerable systems is a gripe in most selected publications, as shown in Table \ref{tab:notification}. It is worth noting that the remediation rates of each selected publication do not directly indicate the efforts and success of the channel selection and messaging infrastructure from the finders. Severity of reported vulnerabilities, impact scale, affected system, and risk management of receiving parties will all contribute to the stakeholders' actions and mitigation progress.
Although 57\% of the vulnerable hosts patched with email notification in \cite{durumeric_heartbleed_2014}, which is relatively higher than other notification operations with remediation rates from lower than 2.25\% to up to around 30\% \cite{cetin_make_2017,stock_hear_2018,stock_hey_2016,li_youve_2016,chen_cvd_2024,stivala_pdf_2024}. Durumeric et al. \cite{durumeric_heartbleed_2014} initiated the notification of the Heartbleed vulnerability 3 weeks after the notable public disclosure. Besides, they had to drop 56\% of the detected vulnerable hosts due to responsible administrators likely having no access to treat the embedded devices~\cite{durumeric_heartbleed_2014}. On the contrary, with a dedicated communication channel and trustworthy notification organisation (ISP Intervention) as in \cite{cetin_tell_2019,rodriguez_iot_2021}, the remediation rate can be significantly higher for more than 75\%; nonetheless, such a scenario requires selective recipients or extra capacity and won't necessarily fit other end-user notifications~\cite{cetin_tell_2019}. Moreover, not every disclosure operation can measure the remediation rate due to the nature of the vulnerabilities and affected parties~\cite{boucher_trojan_2022,koen_rpkiller_2023,moura_stressful_2023,vanhoef_key_2017}. However, Beitis and Vanhoef \cite{beitis_haunted_2025} include both disclosure and notification operations with a measurable impact scale. They did not provide a remediation rate since their work was still a work in progress. Their result is worth observing in the near future.

The lack of best practices in large-scale vulnerability disclosure, particularly in notification, among academic researchers and practitioners, has led to the aforementioned struggles and challenges at each stage, as noted in the title of \cite{moura_stressful_2023}, "Vulnerability Disclosure Considered Stressful". The practice gaps motivated the finders in the selected publications to adopt existing guidelines, reflect on real operations, and contribute to best practices. However, the stress and frustration of the finders deserve attention. Whether it is large-scale disclosure or notification, various vulnerability platforms~\cite{boucher_trojan_2022}, tracking systems~\cite{moura_stressful_2023}, and ticketing systems~\cite{li_youve_2016,cetin_make_2017,chen_cvd_2024} have increased the workload of finders to deliver the messages to the responsible parties in the complex multi-party scenarios. The inaccuracy of existing abuse and generic contact information has caused false positives in contact retrieval and information leaks to unintended recipients, which is mentioned in nearly all selected publications using email as a communication channel. Besides, despite support from national CSIRTs, finders in \cite{vanhoef_key_2017,koen_rpkiller_2023} still encountered the situation that certain stakeholders did not comply with the CVD as best practice in the first place. The feedback from stakeholders is also not always friendly, either in public or private discussions, as reported in \cite{vanhoef_krack_2017,moura_stressful_2023,koen_rpkiller_2023}. In certain cases, to perform timely disclosure to stakeholders, finders still need to put in extra efforts in contacting the vendors directly despite having support organisations \cite{koen_rpkiller_2023} or disclosure programs presented by stakeholders \cite{boucher_trojan_2022}. 

Furthermore, reviewing and responding to stakeholders can take time and effort for the finders. With the ticketing systems as the common practices from stakeholders, the automatic responses in large-scale notifications can result in a high amount of message content to be examined, which is documented in \cite{cetin_make_2017,li_youve_2016,stock_hear_2018,stock_hey_2016,chen_cvd_2024}. Even though automatic messages may share patterns to be categorised, the mixture of multiple languages in messages \cite{li_youve_2016,boucher_trojan_2022,chen_cvd_2024,cetin_make_2017}, unclear stakeholder disclosure policies \cite{chen_cvd_2024,stivala_pdf_2024}, and stakeholders' communication systems requiring manual efforts to register and input messages \cite{cetin_make_2017,chen_cvd_2024} may bring higher than expected workload to finders. This can hinder the effectiveness of large-scale disclosure and notification, and more importantly, the incentives of finders, as discussed in \cite{cetin_make_2017,boucher_trojan_2022,moura_stressful_2023,chen_cvd_2024}. 

To sum up, we conclude by drawing insights from our operations stage model, which is extracted from the selected publications and the experiences shared by support organisations and other communities. This also fulfils our third research question presented in Section \ref{sec:rq}. 
Furthermore, the insights have highlighted current gaps in best practices for large-scale vulnerability disclosure, particularly in notification, across both the academic security research community and other communities.
\section{Best Practices for Large-Scale Vulnerability Disclosure and Notification}
\label{sec:bestpractices}


In this section, we aim to examine the gap in current guidelines and propose new best practices for large-scale vulnerability disclosure and notification based on what we have learned from selected academic publications, communities, and support organisations. The main focus is to help the finders from communities that include academic security researchers, practitioners, and ethical hackers. Yet, the best practices are not limited to finders but also stakeholders to improve the receiving strategies to handle vulnerability reports. We follow the same structure as in the previous section to look into limitations and opportunities in the stages of disclosure and notification operations. In each stage, we discuss common pitfalls to avoid, tradeoffs in method selection, and provide suggestions to the finder and other stakeholders on adopting our best practices.
 

\subsection{Pre-Assessment}
Understanding the impact scale, vulnerability characteristics, vulnerability disclosure or notification, and potential stakeholders is essential for a finder to set up a disclosure or notification operation. Whether it's vulnerability disclosure or vulnerability notification to vendors or to end-users will lead to different communication channels, messaging infrastructure, disclosure policies, and message content. From what we observed and as revealed in the reflections of selected publications, due to the lack of large-scale vulnerability disclosure and notification best practices for different communities, a finder may not fully understand efforts and tradeoffs to conduct the disclosure during the pre-assessment stage, which are documented in \cite{boucher_trojan_2022,moura_stressful_2023}. We list the points below to highlight the common pitfalls and our suggestions:
\begin{itemize}
    \item Security researchers, ethical hackers, and practitioners may not be aware of the difference between disclosure to vendors or notification to end-users, prioritisation of the contact list, and the coverage of national CSIRT support. They may then face stress or frustration during the disclosure operations and struggle with unexpected challenges, as noted by Moura and Heidemann \cite{moura_stressful_2023}. 
    \item A finder and a receiving stakeholder may have different risk assessment standards, resulting in different definitions of vulnerability severity from both ends~\cite{samba_badlock_2016,microsoft_hypedsamba_2016}. As revealed in \cite{li_youve_2016,durumeric_heartbleed_2014}, the receiving parties may set a lower priority or lack the capacity to mitigate the issues earlier than the finders expect. 
    \item We recommend timely consultation with CSIRTs or equivalent support organisations. This can help a finder comprehend the potential impact of vulnerability and the scale of the notification to stakeholders, as revealed in the timeline by van Hove et al. \cite{koen_rpkiller_2023} and suggested by support organisations~\cite{first_guideline_2020,oecd_product_2021, ncsc_guideline_2018,householder_cert_2017, BOD_CISA}. Moreover, this can also help a finder estimate potential stakeholder responses and prepare the message handling in advance. 
\end{itemize}
The next common question is which combination of channel, messaging infrastructure, disclosure policy and interaction can be the most effective in each case. These aspects will be discussed in the following stages.

\subsection{Communication Channel}
Choosing the most effective channels to perform disclosure is the key to reaching out to stakeholders \cite{wogalter_chip_1999}. We provide our suggestions on communication channels for large-scale vulnerability disclosure and notification separately.

In the case of \textbf{vulnerability disclosure to vendors}, there are already comprehensive guidelines and support mechanisms in place to help finders inform the receiving stakeholders:
\begin{itemize}
    \item A finder can seek vulnerability disclosure policies or programs of stakeholders \cite{boucher_trojan_2022,moura_stressful_2023}, which are possibly indicated by stakeholders providing a security.txt~\cite{rfc9116}.
    \item  A finder can also look for bug bounty programs or vulnerability platforms to issue the vulnerability report. However, stakeholders may also have outsourced such channels to third-party platforms such as HackerOne and Bugcrowd~\cite{walshe_bugbounty_2020}, and having a legal agreement in disclosure~\cite{macnish_ethics_2020}. 
    \item A finder can reach out to national CSIRTs to consult on a possible communication channel~\cite{vanhoef_key_2017,vanhoef_krack_2017}, if the pre-assessment of potential stakeholders or the impact of a vulnerability is unclear. National CSIRTs may provide notification services, coordination support or platforms such as VINCE by CERT/CC~\cite{vince} to efficiently identify and inform vendors~\cite{first_guideline_2020,ncsc_guideline_2018}.
\end{itemize}

In the case of \textbf{vulnerability notification to end-users}, finding an effective and efficient communication channel still remains a significant challenge despite the current guidelines:
\begin{itemize}
    \item A finder can first consider selecting known stakeholders with clear contact information~\cite{moura_stressful_2023,beitis_haunted_2025}, then seek the rest of the stakeholders' contact information.
    \item  A finder should then be mindful of the messaging prioritisation and scheduling if the response time from the initial list of stakeholders takes longer than expected~\cite{moura_stressful_2023}.
    \item A finder should be aware that delegation to national CSIRTs mostly works but may not be effective in every scenario, as the interests, notification coverage, and capacity of different national CSIRTs or equivalent support organisations may vary \cite{durumeric_heartbleed_2014,cetin_make_2017,stock_hear_2018}.
    \item A finder should consider reaching out to key communities or platforms that support remediation tracking \cite{boucher_trojan_2022, moura_stressful_2023}, depending on the affected parties. Phone calls and posts are more of an option when stakeholders suggest doing so or the regimes have such practice \cite{maass_letter_2021, stock_hear_2018}.
    \item We do not recommend using email to perform large-scale vulnerability notifications to end-users, even though this has been the most commonly used option. Setting up an email infrastructure for doing large-scale notifications is difficult, keeping in mind all of the spam prevention tools currently in use.
    \item We recommend the use of The Shadowserver Foundation as a notification channel, as implemented by Beitis and Vanhoef \cite{beitis_haunted_2025}. The organisation can perform active scans and has trusting relationships with key stakeholders that support the notification infrastructure. This has the added advantage that end-users are not overwhelmed with notification campaigns from different finders.
\end{itemize}

Email as a communication channel in large-scale notifications faces the additional challenge of finding contact information for systems on the Internet. The accuracy of the contact information presented on web pages is not guaranteed~\cite{maass_letter_2021}. Finding contact information for IP addresses is notoriously hard. Although WHOIS and RDAP are methods to retrieve the contact information, the current results of both are often not accurate~\cite{fernandez_whois_2024}. Moreover, the contact information available is often meant for abuse notifications, not for vulnerability notifications. While the `security.txt' standard\cite{rfc9116} works for websites, there is no such alternative for IP addresses.

\subsection{Messaging Infrastructure}
The messaging infrastructure will depend on the selected communication channels. Finders should follow the indicated preference of vendors in using the contact method, usually email, or messaging platforms such as bug bounty programs, VINCE, or other vulnerability report platforms. It should be noted that the effectiveness of large-scale email notifications leaves a lot to be desired, as can be seen in Table~\ref{tab:notification}. \textbf{If an email infrastructure is still used:}
\begin{itemize}
    \item We recommend that the sending email address be from a known domain name or with an organisation to increase the delivery rate~\cite{max_cvd_2023,stock_hear_2018,chen_cvd_2024}. One should also be aware of the implementation of their mail service; with the mail service outsourcing trend, there can be rate limits and extra policies to examine beforehand~\cite{chen_cvd_2024}.
    \item A finder should be aware that recipients may send automatic responses, divert to ticketing systems, internal communication or management systems, request feedback forms~\cite{cetin_make_2017,durumeric_heartbleed_2014,chen_cvd_2024,stock_hear_2018}, or outsourced platforms with different disclosure policies~\cite{boucher_trojan_2022,koen_rpkiller_2023}.
\end{itemize}

The scalability of the messaging infrastructure is the biggest challenge when performing large-scale vulnerability notifications. Although organisations such as CERT/CC with VINCE and national CSIRTs with vulnerability notification systems can help with large-scale disclosure, there is still no optimised channel and infrastructure for large-scale vulnerability notifications in the existing guidelines \cite{ImprovingPrivateSector}.
\textbf{Nonetherless, aside from the email infrastructure:}
\begin{itemize}
    \item A finder can provide web pages to describe the intention of the disclosure, vulnerability information, remediation, and disclosure policy. This can help reduce the content in email messages and provide a static source for stakeholders to help track the issues in the long term~\cite{vanhoef_krack_2017,cetin_make_2017,stock_hear_2018,chen_cvd_2024}.
    \item We recommend checking with a support organisation or partner for messaging infrastructure. National CSIRTs, PSIRTs, ISPs, and more stakeholders have been improving external or internal notification mechanisms \cite{InformationSecurityEarly_Japan, cetin_tell_2019, rodriguez_iot_2021}. 
    \item We recommend The Shadowserver Foundation, with its established infrastructure and notification experiences, as an effective option for network scanning, identifying vulnerable systems, and reaching affected parties \cite{ShadowserverFoundation}. 
\end{itemize}

In addition, it is worth noting that the adoption of bug bounty programs and vulnerability report platforms as accepted best practices among stakeholders has been growing over the years. However, while well-established stakeholders may have mature report-handling policies and support mechanisms to receive vulnerability reports at scale \cite{boucher_trojan_2022}, other stakeholders may still lack the capacity and experience to present a clear disclosure policy or program in their public information or ticketing system \cite{chen_cvd_2024}. Although some regimes with laws and regulations \cite{cra, BOD_CISA, ncsc_guideline_2018} that mandate stakeholders to take action upon notification have successful cases, such as \cite{maass_letter_2021}, it is not guaranteed that stakeholders will initiate mitigation, particularly in multi-party notifications at the Internet-wide scale with diverse stakeholders involved, as presented in \cite{max_cvd_2023, chen_cvd_2024}.
Moreover, even vendors or other stakeholders may still struggle to notify their clients and ensure trust from the receiving end-users with correct contact information  \cite{ImprovingPrivateSector}. Eventually, it will be easier for finders to perform vulnerability disclosure and notification at scale once the adoption of the practices gets higher.

\subsection{Disclosure Policy and Message}

\textbf{Composing a disclosure message} is never trivial. Among stakeholders, a finder and affected parties can have different preferences and procedures for handling the messages~\cite{walshe_coordinated_2022,max_cvd_2023}. 
\begin{itemize}
    \item In vulnerability disclosure, stakeholders with bounty programs or disclosure policies may provide clear instructions or at least contact information to initiate the disclosure~\cite{walshe_bugbounty_2020}. In contrast, recipients of the vulnerability notification may not provide enough information on how they will handle a disclosure message.
    \item A finder should know that stakeholders like network operators may prefer extensive information that includes more details and remediation~\cite{max_cvd_2023,ncsc_guideline_2018}, given that the affected party follows a certain time constraint policy on remediation~\cite{chen_cvd_2024}. Meanwhile, stakeholders like cloud providers or domain owners may forward the message to their clients with limited communications~\cite{chen_cvd_2024,li_youve_2016}. 
    \item A finder should be aware that not every stakeholder would accept longer messages or attachments regarding the mail server filter, ticketing systems, and spam filter~\cite{cetin_make_2017, stock_hear_2018, chen_cvd_2024}.
    \item We recommend that a finder ensure that the initial message remains brief and does not necessarily reveal every detail of the vulnerability in case of an information leak or legal action~\cite{vanderham_cvd_2023, max_cvd_2023}.
\end{itemize}

\textbf{In conversations with stakeholders} after initiating vulnerability disclosure or notification.
\begin{itemize}
    \item A finder should consider that large-scale vulnerability disclosure and notification may get manual or automatic responses from various communication systems in different languages. In most cases, we have seen English used in the responses; other languages are used as support~\cite{durumeric_heartbleed_2014,li_youve_2016,chen_cvd_2024}. 
    \item A finder should be aware that messages in different languages may increase the processing time and cause confusion, especially if disclosure policies are presented in non-native languages to a finder or a receiving party.
    \item  A finder may need to put extra effort into reformulating the messages based on the limitations of the disclosure form or editors in the provided text or system when using stakeholders' communication systems \cite{chen_cvd_2024}.
\end{itemize}

\textbf{To inform the intended disclosure or notification}, it's important to provide the motivation, mitigation scheduling, and legal terms, if possible. As we observed in the selected publications, disclosure policies and equivalent documentation are not widely implemented. This has resulted in both finders and receiving parties having inefficient communication.
\begin{itemize}
    \item A finder should be aware of the stance of stakeholders during disclosure or notification. Not every stakeholder may want to adopt the best practices for different reasons. This has been reported in two disclosure operations~\cite{vanhoef_key_2017,koen_rpkiller_2023} with vendors refusing to participate in the remediation schedule despite having a national CSIRT as coordinator. Still, we encourage stakeholders to adopt best practices and participate in disclosure or notification operations.
    \item We recommend that a finder and support organisations establish an outgoing disclosure policy~\cite{vanderham_cvd_2023,reidsma_operationalizing_2023, ncsc_guideline_2018}, which includes legal terms, disclosure schedules, message templates, or exemptions, as implemented in~\cite{stivala_pdf_2024, koen_rpkiller_2023, chen_cvd_2024}. This allows recipients to understand the disclosure procedure from a finder and protects the finder from unwanted behaviours, such as legal actions \cite{macnish_ethics_2020} and public criticism \cite{koen_rpkiller_2023}. 
    \item We recommend that the stakeholders include a disclosure policy\cite{householder_cert_2017,ncsc_guideline_2018,vanderham_cvd_2023} or security.txt\cite{rfc9116} in their disclosure programs or responses. This helps a finder to initiate the disclosure and notification operations with the stakeholders with better preparation.
\end{itemize}

\subsection{Post-Review}
With the fast-changing nature of the network landscape and the growing number of vulnerabilities, it is important to have best practices up to date and mitigate security issues in time. We have seen publications focusing on the exploits, attacks, and network traffic before and after the public disclosure. However, documentation focusing on large-scale vulnerability disclosure and notification is relatively scarce in certain communities, as noted in \cite{chen_cvd_2024,moura_stressful_2023, stock_hear_2018}. Nevertheless, we have seen that finders in \cite{chen_cvd_2024, stivala_pdf_2024,beitis_haunted_2025} could learn from prior best practices and contribute to large-scale notification. Handling disclosure or notification procedures can be a nerve-racking trial, as revealed in nearly all selected publications. It is crucial to have evolving best practices to help a finder prepare for challenges and ease the stress during operations.
\begin{itemize}
    \item We recommend that security researchers, ethical hackers, practitioners, and more finders document their disclosure or notification operations with their experiences, considerations, and outcomes.
    \item We recommend that finders review the impact before and after vulnerability disclosure or notification, which can help finders reflect on their progress and improve best practices.
    \item A finder can record the challenges and stress encountered during a disclosure or notification operation. The tradeoffs and considerations in different scenarios are also worth recording.
    \item A finder can provide a timeline as a figure, like in \cite{boucher_trojan_2022,koen_rpkiller_2023,moura_stressful_2023,durumeric_heartbleed_2014,stock_hear_2018,stock_hey_2016} or as text, like in \cite{chen_cvd_2024,stivala_pdf_2024}, for each disclosure or notification stage, which can help understand the operation development and its impact over time. 
    \item A finder can track the number of remaining vulnerable systems before and after disclosure or notification, if the vulnerable systems can be traced through network scanning, stakeholder engagement, or user communication. 
    \item A finder can track the mitigation progress from weeks~\cite{durumeric_heartbleed_2014}, months~\cite{cetin_tell_2019,chen_cvd_2024}, or more than a year~\cite{stivala_pdf_2024} if the situation permits. The tracking update can be presented as a webpage~\cite{vanhoef_key_2017,boucher_webtrojan_2022} , experience reports \cite{moura_stressful_2023}, or follow-up academic publications~\cite{durumeric_heartbleed_2014,stivala_pdf_2024,boucher_trojan_2022,stock_hear_2018,vanhoef_kraken_2018} to present disclosure or notification updates.
\end{itemize}


Last but not least, as we have observed in the academic security research community, academic researchers as finders may show different preferences and limitations compared to bug bounty hunters or vendors as finders. Researchers may need to handle the publication cycle aside from the disclosure or notification procedure, which brings extra time constraints and stress in their work. Further, we have observed that not every academic researcher has the capacity and prior experience to conduct disclosure or notification operations and finish the documentation within the academic publication period. We believe such a situation can arise in different forms for finders in some other communities. With the increasing adoption of support mechanisms such as the disclosure and bug bounty program, the stress of conducting vulnerability notification at scale can be mitigated, but not greatly reduced, as discussed in Section \ref{sec:lessons-postreview}. This is the main motivation for proposing best practices with suggestions in each operation stage to help finders, considering scalability, effective communication, and finder protection.

We have answered our final research question presented in Section \ref{sec:rq}. However, more perspectives on vulnerability disclosure and notification from different finders and stakeholders also require attention. We hope more finders and other stakeholders can benefit from our recommendations and contribute to the security communities by documenting their own experiences and improving best practices.

\section{Conclusion}

The practice of doing academic research on vulnerabilities is growing in popularity. Even though we have best practices for vulnerability disclosure, this does require more attention when this scales up. We note that there is a difference in practice between vulnerability disclosure and vulnerability notification, especially with regard to the stakeholders involved. We have analysed trends in academic work and the security community, and propose new best practices to bridge the gap between the existing guidelines and the limitations in actual operations.  We believe that our best practices give researchers, ethical hackers, and practitioners a clear direction to inform stakeholders at scale with less friction.  With our suggestions, stakeholders can prepare for the disclosure or notification message response and mitigation. We encourage all the stakeholders, including finders, vendors, and end-users, to not only bring in the best practices but also document and publish their experiences to help improve future disclosure and notification operations.

\section{Acknowledgments}

Blanked for review.

\bibliographystyle{ACM-Reference-Format}
\bibliography{reference}

\appendix









\end{document}